\documentclass[12pt]{article}
\begin{document}

\title{\bf The Momentum Four-Vector in the $e\psi$N Formalism and the Angular Momentum
Imparted to Test Particles by Gravitational Waves}
\author{Asghar Qadir\thanks{Senior Associate, Abdus Salam International Center for
Theoretical Physics, Trieste, Itlay}, M. Sharif\thanks{Permanent
address: Department of Mathematics, Punjab University, Lahore-Pakistan.}
and M. Shoaib\\
Department of Mathematics, Quaid-i-Azam University,\\ Islamabad,
Pakistan}

\date{}
\maketitle
\begin{abstract}
Since gravitational waves are solutions of Einstein's field
equations with a zero stress-energy tensor, the reality of these
waves was questioned. To demonstrate it, the momentum imparted to
test particles by such waves was evaluated. A closed form expression
for this quantity was provided by Qadir and Sharif, using an
extension of the pseudo-Newtonian formalism. That formalism carried
with it the zero component of the momentum vector, which could not
be interpreted as the energy imparted to the test particle. Sharif
proposed that it may represent the angular momentum imparted to test
particles by gravitational waves. In this paper it is shown that
this interpretation is not valid. An alternative explanation has
been provided.
\end{abstract}

04.20.Cv-Fundamental problems and general formalism,\\
04.30 - Gravitational waves: theory.

\section{Introduction}

Einstein's General Relativity (GR) predicts the existence of
gravitational waves as solutions of the vacuum Einstein equations
[1]. The significance of these solutions as gravitational waves was
questioned on account of the fact that they cannot posses any
energy, since the stress-energy tensor is zero The problem was
resolved by Ehlers and Kundt [2], Pirani [3] and Weber and Wheeler
[4], by considering test particles in the path of the waves. It was
shown that these particles acquired momentum from the waves. The
standard procedure [5] to evaluate the momentum is fairly
complicated. Qadir and Sharif [6] provided a closed form expression
for the momentum imparted in arbitrary spacetimes, using an
extension [7] of the pseudo-Newtonian ($\psi$N) formalism [8]. The
original formalism applied to only static spacetimes and, in the
appropriate Fermi Walker frame [9], gave a purely spatial force
vector. Its extension to non-static spacetimes introduced a zero
component for the force vector. The problem was to interpret this
extra term. The original $\psi$N force could be expressed as the
gradient of a scalar potential but the spatial part of the $e\psi$N
force had to be written as the spatial gradient of one potential and
the zero component as the time derivative of another. This was not
satisfactory.The alternative method [6] of treating the force
4-vector as the proper time derivative of the momentum 4-vector was
adopted. This procedure provided the momentum imparted to test
particles by gravitational waves as
\begin{equation}
p_{\mu}=\int F_{\mu}dt,
\end{equation}
where
\begin{equation}
F_{0}=m[(\ln(\frac{A}{\sqrt{g_{00}}}))_{,0}-\frac{g^{ij}_{,0}g_{ij,0}}{4A}],
\end{equation}
\begin{equation}
F_{i}=m(\ln\sqrt{g_{00}})_{,i},
\end{equation}
\begin{equation}
A=(\sqrt{-\det(g_{ij}}))_{,0}.
\end{equation}
In this paper the problem of identification of the zero component of
the momentum 4-vector was mentioned. Sharif [10] proposed that it
could be interpreted as the spin imparted to test rods. In the next
section this suggestion will be reviewed and it will be demonstrated
that it does not work. To find an alternative check on its validity,
the geodesic analysis [2] for the angular momentum imparted to test
articles by gravitational waves is undertaken in sect. \textbf{3}.
Finally, in the conclusion a proposal for the interpretation of the
zero component of the momentum 4-vector is discussed.

\section{Spin Imparted to Test Particles by Gravitational Waves}

Sharif [10] considered a test rod of length $\lambda$ in the path of
a gravitational wave whose preferred direction is given by $l^{i}$
in the preferred reference frame. It was argued that the maximum
angular momentum imparted would be
\begin{equation}
S^{i}=p_{0}l^{i},
\end{equation}
where $l^{i}$ is the vector representing the rod. Thus the maximum
angular momentum imparted to a test rod, when it lies in the plane
perpendicular to the preferred direction, is
\begin{equation}
S=p_{0}\lambda=m \lambda
\int[(\ln(Af)_{,0}-g^{ij}_{,0}g_{ij,0}/4A]dt.
\end{equation}
Hence the physical significance of the zero component of the
momentum 4-vector would be that it provides an expression for the
spin imparted to a test rod in an arbitrary spacetime.

This formula was applied to plane and cylindrical gravitational
waves to give, respectively, a constant $S$ and
\begin{equation}
S=-m\lambda[({1+AJ}_{0}/ \omega \rho J'_0)\ln |1 - 2A \omega \rho
J'_0 \cos(\omega t) + constant.
\end{equation}

Notice that here can be no spin angular momentum imparted to test
particles in a perfectly homogeneous and isotropic cosmological
model; its high degree of symmetry, in particular, spherical
symmetry - is incompatible with spin being imparted to particles.
However when we use the above formula for cosmological models, it
gives exactly this error [11]. For example, consider the Friedmann
model, which is isotropic and homogeneous. For the flat Friedmann
model
\begin{equation}
S=m\lambda\ln \frac{\eta}{2}.
\end{equation}
For the closed Friedmann model there was a problem, as a straight
forward application of the formula led to an infinite value of
$F^{0}$ at the phase of maximum expansion. This arose because the
coefficient of $F^{0}$ in the constraint equation vanished. The
problem was resolved by re-setting the zero of acceleration at the
phase of maximum expansion. Setting $S$ to zero at the phase of
maximum expansion.
\begin{equation}
S=m\lambda[\ln\sqrt{1-\cos \eta} - \frac{3}{8} \cos \eta
+\frac{1}{16} \cos^{2}\eta+\ln\sqrt{2}+\frac{5}{16}].
\end{equation}
For the open Friedmann model
\begin{equation}
S=-m\lambda\ln\frac{\sinh\eta}{\cosh\eta-1}.
\end{equation}

Equations (8), (9) and (10) give a non-zero spin imparted to test
particles by flat, open, and closed Friedmann models. Since no spin
can be imparted, the spin interpretation of $p_{0}$ cannot be
correct.

For the De Sitter universe in usual coordinates the spin turns out
to be constant and in the Lemaitre form to be
\begin{equation}
S=-m\lambda(\Lambda/3)^{1/2}t+constant.
\end{equation}
This does not seem reasonable, because for the former version we
could get rid of the spin by choosing the constant to be zero, while
for the latter we cannot. Thus it is clear that the interpretation
is not even internally consistent.

\section{The Geodesic Analysis for the Angular Momentum
Imparted to test Particles in Various Cosmological Models}

Consider a time-like congruence of world lines ( not necessarily
geodetic) with tangent vector $u^{a}$. Decompose $u_{a;b}$ by means
of the operator $h_{ab}$ projecting into the infinitesimal 3-space
orthogonal to $u^{a}$ [2]:
\begin{equation}
u_{a;b}=-\omega_{ab}+\sigma_{ab}+\frac{1}{3}\vartheta
h_{ab}-\dot{u}_{a}u_{b},
\end{equation}
where
\begin{equation}
-\omega_{ab}\equiv
u_{[a,b]}+\dot{u}_{[a}u_{b]},\quad\dot{u}_{a}=u_{a;b}u^{b},
\end{equation}
\begin{equation}
\vartheta \equiv u^{a}_{;a},\quad h_{ab} = g_{ab} +u_{a}u_{b},
\end{equation}
\begin{equation}
\sigma_{ab}\equiv u_{(a;b)}+\dot{u}_{(a}u_{b)}-\frac{1}{3} \vartheta
h_{ab}\quad(\sigma^{a}_{a}=0).
\end{equation}

For an observer along one of the world lines and using Fermi
propagated axes, $\omega_{ab}$ describes the velocity of rotation,
$\sigma _{ab}$ shear and $\vartheta$ the expansion of the cloud of
neighboring free particles. Since the $\psi$N-formalism uses the
Fermi-Walker frame, it could be expected that the results of this
analysis should be consistent with it, as indeed they are for the
momentum imparted. For our purpose only $\omega_{ab}$ is needed.
Choose the coordinates so that the tangent vector is
$u^a=\frac{1}{\sqrt{g_{00}}}\delta^{a}_{0}$. Thus, from eq.(13) we
have
\begin{eqnarray} \omega_{i0}=-\frac{1}{2} \left\{
\begin{array}{ccc}
   & 0 &  \\
  0 &  & i
\end{array}
\right\} \quad(i=1,2,3),
\end{eqnarray}
for the cases when $g_{00}=1$. For the Friedmann or the Lemaitre
form of the De Sitter universe this gives zero. Even for the De
Sitter universe in usual coordinates we get a zero spin, as
required. For the G$\ddot{o}$del universe the spin turns out to be
the physically reasonable.
\begin{equation}
\omega_{10}=-\frac{1}{4}
\frac{am^{2}(x)}{l(x)+m^{2}(x)},\quad\omega_{20}=\omega_{30}=0.
\end{equation}

For gravitational waves we have to introduce a procedure so that the
spacetime is Minkowski before the wave arrives and acquires a
non-static metric afterwards. The geodesic analysis for
gravitational waves is given elsewhere [12] as it is not directly
relevant here. The expressions are quite complicated, involving
delta functions so as to incorporate the step function on the
metric. They are consequently omitted here.

\section{Conclusion}

In the absence of any momentum 3-vector and $F_{0}$, we should be
able to identify $p_{0}$ with $E$. Thus we define
\begin{equation}
\Delta E=p^{0}-E=g^{00} \int
F_{0}dt-[(m^2-p^{i}p^{j}g_{ij})^{\frac{1}{2}}+m],
\end{equation}
where the last term corresponds to an integration constant. Let us
calculate this difference for the three cases of the Friedmann
metric. They give $\Delta E=S/\lambda$, with $S$ given by eqs.(8),
(9) and (10) for the corresponding cases.

The three expressions have the same asymptotic behavior for
sufficiently small values of $\eta$, namely $\Delta E \sim
m\ln\eta$. At the phase of maximum expansion of the closed model we
get $\Delta E= m(\ln2+3/4)$. We could equally well have set $\Delta
E=0$ at the phase of maximum expansion and had a difference for it
from the other two cases for small $\eta$ (in the constant term).
Note that for all the three models $\Delta E$ diverges as $\eta$
goes to zero and it also diverges as $\eta$ goes to $2\pi$ for the
closed model. For $k=+1$ we have chosen to display the constant term
so $\Delta E$ for the closed Friedmann model with $m$ taken to be
unity. This becomes infinite at the bang and the crunch and becomes
zero at the phase of maximum expansion. $\Delta E$ for the flat
Friedmann model, again $m$ is taken to be unity. $\Delta E$ diverges
at the bang only. $\Delta E$ for the open Friedmann model, again $m$
is taken to be unity. Again, $\Delta E$ diverges at the bang only.
that $\Delta E=0$ at $\eta=\pi$. The proposal does not seem
inconsistent, but still needs further discussion. For this purpose
we shall briefly recall the key points of the definition of the
$e\psi$N-force.

The original $\psi$N-formalism, constructed for static spacetimes
only, was based on the observation that even in free-fall the tidal
force can be operationally determined by an accelerometer consisting
of two masses connected by a spring, whose end can move as a needle
on a dial. Using Riemann normal coordinates, the extremal value of
this tidal force gives the directional derivative of the
relativistic analogue of the Newtonian gravitational force, called
the $\psi$N-force. The extension to non-static spacetimes was
achieved by retaining the requirement of Riemann normal coordinates
spatially, but dropping it temporally. Consequently, a "memory" of
the zero-setting of the accelerometer is built into the formalism.
Note that in the case of the closed Friedmann model the "memory"
comes from the phase of maximum expansion rather than from the bang,
or any other time prior to the present.

The fact that $F_{0}$ has the zero setting built into it is the key
to the interpretation of both $F_{0}$ and $p_{0}$. Essentially, as
$F_{0}$ has the dimensions of power, it should correspond to the
"power" imparted to the test particle, or absorbed from it, by the
gravitational field of the spacetime. For a static spacetime this
would be zero, as energy would be conserved. However, in a
time-varying spacetime there must be energy imparted to, or
withdrawn from, all (test) particles in it. This would provide a
direct measure of the extent of non-conservation of energy in the
spacetime.

The interpretation of $p_{0}$ in the $\psi$N-formalism is now clear.
It is the \emph{extra} energy imparted to test particles or
extracted from them) by the gravitational field, in the following
sense. The energy imparted, as at the time considered, relative to
the spacetime then is the usual.
\begin{equation}
E_{T}=E+\int F_{0}dt=E+p_{0}.
\end{equation}
It should be borne in mind that $p_{0}$ can be negative. It may
also be remarked that the previously introduced "second
potential", $U$, is the same as $p_{0}$ (up to an arbitrary
constant of integration, not fixed for either of them).

The question still unanswered is how to convert this \emph{measure}
of the gravitational energy of the field into an actual expression
for it, without reference to the test particle. The problem is that
the mass of the test particle enters into the expression for the
energy. The problem is that the mass of the test particle enters
into the expression for the energy. One would be tempted to replace
it by $\rho dV,$ and use the energy density in the spacetime to
obtain $\rho$. This procedure cannot work. To see this, consider a
gravitational wave spacetime. Clearly, it does not have any
stress-energy tensor, and hence one would have to put $\rho=0$.
However, there is obviously a gravitational energy in the field,
$p_{0}\neq0$, in general.

One needs to, somehow, use the very gravitational energy of the
field we are trying to evaluate to give the required $\rho$. How
this is to be done is not clear at present.

\vspace{0.5cm}

{\bf Acknowledgments}

\vspace{0.5cm}

One of us (AQ) would like to thank Prof. R. Ruffini for financial
support at ICRA, where some of this work was done and two of us (AQ
and M. Shoaib) would like to acknowledge the Quaid-i-Azam University
Research Fund for support facilities for this work.

\vspace{0.5cm}

{\bf References}

\begin{description}

\item{[1]} Misner C.W., Thorne K.S. and Wheeler J.A.,
\emph{Gravitation} (W.H. Freeman, San Francisco) 1973.

\item{[2]} Ehlers J. and Kundt W., \emph{Gravitation: An Introduction to
Current Research}, edited by L. Witten (Wiley, New York) 1962, p.
49.

\item{[3]} Pirani F.A.E., \emph{Gravitation:An Inroduction to Current
Research}, eidted by L. Witten (Wiley, New York) 1962, p. 199.

\item{[4]} Weber J. and Wheeler J.A., Rev Mod. Phys., \textbf{29}(1957)509.

\item{[5]} Weber J., \emph{General Relativity and Gravitational Waves}
(Interscience, New York) 1961.

\item{[6]} Qadir A. and Sharif M., Phys. Lett. \textbf{A167}(1992)331.

\item{[7]} Qadir A. and Sharif M., Nuovo Cimento
\textbf{B107}(1992)1071;\\
Sharif M., Ph.D Thesis, Quaid-i-Azam University (1991).

\item{[8]} Qadir A., Nuovo Cimento \textbf{B112}(1997)485;\\
Mahajan S.M., Qadir A. and Valanju P.M., Nuovo Cimento
\textbf{B65}(1981) 404;\\
Quamar J., Ph.D. Thesis, Quaid-i-Azam University (1984);\\
Qadir A. and Quamar J., \emph{Proceedings of the Third Marcel
Grossman Meeting on General Relativity}, edited by Hu Ning (Science
Press and North Holland Publishing Co.) 1983, p. 189.

\item{[9]} Qadir A. and Zafarullah I., Nuovo Cimento \textbf{B111}(1996)79.

\item{[10]} Sharif M., Astrophys. Space Sci., \textbf{253}(1997)159.

\item{[11]} Sharif M., Astrophys. Space Sci., \textbf{262} (1999)297.

\item{[12]} Shoaib M., M. Phil dissertation, Quaid-i-Azam
University (1999).
\end{description}
\end{document}